\documentclass[12pt]{article}

\usepackage{graphicx}

\usepackage{amssymb}
\usepackage{color}
\newcommand{\nc}{\newcommand}

\nc{\lm}{\lambda}
\nc{\ve}{\varepsilon}
\nc{\vtheta}{\vartheta}
\nc{\bu}{\mathbf{u}}
\nc{\bd}{\mathbf{d}}
\nc{\red}{\color{red}}
\nc{\blu}{\color{blue}}
\nc{\bl}{\color{green}}

\nc{\ZN}{{\mathbb Z}_N}
\nc{\CH}{{\cal H}}
\nc{\ks}{{\sf k}'}

\nc{\SIQC}{{SCPC}}
\def\r#1{(\ref{#1})}
\nc{\be}{\begin{equation}} \nc{\ee}{\end{equation}}
\nc{\bea}{\begin{eqnarray}} \nc{\eea}{\end{eqnarray}}
\let\ds=\displaystyle
\nc{\ny}{\nonumber}

\begin{document}

\begin{center}{\bf \LARGE Spin operator matrix elements\\ in the superintegrable chiral Potts\\ quantum chain}
\end{center}

\bigskip
\begin{center}
{\bf N.~Iorgov, V.~Shadura, Yu.~Tykhyy}\\ {\small
              Bogolyubov Institute for Theoretical Physics, Kiev 03143, Ukraine \\
              {\tt iorgov@bitp.kiev.ua}}\\[5mm]
           {\bf S.~Pakuliak}\\{\small
              Bogoliubov Laboratory of Theoretical Physics,
              Joint Institute for Nuclear Research,\\ Dubna 141980, Moscow region, Russia\\
              {\tt pakuliak@theor.jinr.ru}} \\ [5mm]
           {\bf G.~von~Gehlen} \\{\small
           Physikalisches Institut, Universitaet Bonn, Nussallee 12, 53115 Bonn, Germany\\
              {\tt gehlen@th.physik.uni-bonn.de}}
\end{center}

\bigskip

\abstract{
We derive spin operator matrix elements between general eigenstates of the   superintegrable
$\ZN$-symmetric chiral Potts quantum chain of finite length. Our starting point is
the extended Onsager algebra recently proposed by R.Baxter. For each pair of spaces (Onsager sectors)
of the irreducible representations of the Onsager algebra, we calculate
the spin matrix elements between the eigenstates of the Hamiltonian of the quantum chain in factorized form,
up to an overall scalar factor. This factor is known for the ground state Onsager sectors.
For the matrix elements between the ground states of these sectors
we perform the thermodynamic limit and obtain the formula for the order parameters.
For the Ising quantum chain in a transverse field ($N=2$ case) the factorized form for the matrix elements
coincides with the corresponding expressions obtained recently by the Separation of Variables Method.
}


\bigskip

\section{Introduction}
\label{intro}

The solution of the Ising model by Onsager \cite{Ons} in 1944 was a major breakthrough in mathematical physics.
The subsequent calculation of the Ising magnetization by Yang \cite{Yang} in 1952 (after the announcement of the
formula by Onsager \cite{Ons1} in 1949) was an achievement of similar importance. The main tool in Onsager's work \cite{Ons} was the
introduction of the algebra which now carries his name. Later many alternative methods for solving the
Ising model were invented and the interest in Onsager's algebra faded away. It was not revived until 1985 when
the superintegrable chiral Potts quantum chain (\SIQC) was formulated \cite{GR}.
For more than 40 years the Ising model had remained the only known representation of Onsager's algebra.
Then the \SIQC\ provided a whole series of new Onsager algebra representations. In many respects, the \SIQC\
is the ${\mathbb Z}_N$ generalization of the ${\mathbb Z}_2$ Ising model.
Soon, in 1988, Baxter  \cite{BaxFE}, using functional relations of the 2-dimensional integrable chiral Potts model
 \cite{BAYP,YP}
calculated the partition function of the \SIQC. The formula for the order parameter (the generalization of the
famous Onsager--Yang result for the Ising model) was conjectured in 1989 \cite{ACPT} from a perturbative calculation
up to ${{\sf k}'}^5$ for arbitrary $N$ (see also \cite{HL}). However, it took until 2005
that Baxter \cite{BaxOPL,BaxOP} succeeded to prove the conjectured formula. His derivation used functional
relations for the $\tau_2$-model and natural analytical assumptions.

Nevertheless, Baxter \cite{Bax2} insisted that a purely algebraic derivation
of the order parameter
formula would be desirable. So, recently,
he extended the Onsager algebra by a \SIQC-spin operator \cite{Bax3}.
In order to close the algebra one has to include many new operators which
are produced by commutations
with the generators of the Onsager algebra. R.Baxter conjectured that
the two simplest relations in the extended Onsager algebra uniquely define
(up to a common multiplier)
the matrix elements of the spin operator between vectors from the ground
state Onsager sectors.
Baxter suggested that this conjecture will imply the determinant
representation (7.11) in \cite{Bax2}
which in the thermodynamic limit will give the formula for the order
parameters.

Our starting point for this research was that we compared the factorized formula (78) in
\cite{BBS2} for the Ising quantum chain spin operator matrix elements
obtained by the Separation of Variables Method
to Baxter's conjectured equation (3.45) of \cite{Bax3}.
In the case of the Ising chain, after a simple transformation given in Section 4.1 of the present paper,
 (3.45) of \cite{Bax3} should imply (78) of \cite{BBS2}.
Since Baxter's conjecture
was formulated for arbitrary parameters, it was natural to suggest that
the factorized formula (78) in \cite{BBS2} for the ground Onsager sectors
is valid also for the ground Onsager sectors in \SIQC\ if one uses appropriate
parameters labelling these sectors.
Numerical calculations for the finite length chain together
with calculations for the thermodynamic limit (see Appendix C) strongly
supported our conjecture. This encouraged us to prove the Baxter's conjecture
(3.45) from \cite{Bax3}.

In the present paper we generalize this conjecture to arbitrary Onsager sectors and
prove it. We found that the matrix elements of
spin operators between the eigenvectors of the \SIQC\
Hamiltonian for arbitrary Onsager sectors can be presented in a factorized form.
The problem of  finding the matrix elements of the spin operator is reduced
to the problem
of finding non-zero matrix elements of the spin operator between
some particular vectors from the corresponding two irreducible
representations of the Onsager algebra.
In the case of ground state Onsager sectors this
gives an exact factorized formula for the matrix elements of spin operators
between such type eigenvectors. Finally, taking the thermodynamic limit of
ground state matrix elements we derive
the Albertini et al. formula \cite{ACPT,HL} for the order parameters.
We would like to note that factorized formulas for the spin matrix elements
exist also for the 2D Ising model \cite{BL} and for the quantum $XY$-chain \cite{Iorgov}.

Recently, state vectors for the superintegrable chiral Potts model
were investigated exploiting the $sl(2)$-loop algebra
symmetry of the $\tau_2$-model \cite{BS,BBP}, which is closely related to the chiral Potts model \cite{NDCP,ND,AuYP1,AuYP2}.
Since our approach avoids direct use of the relation to the $\tau_2$-model, we shall not go into details
of this exciting work.

This paper is organized as follows: In Section 2 we define the \SIQC\ by its Hamiltonian and use the fact that its two
parts generate the Onsager algebra and imply the Ising form of the eigenvalues. Baxter's extension and the
basic two resulting equations for the spin operator are stated. In Section 3 we develop the consequences of
Baxter's two equations for spin matrix elements: including higher Onsager sectors and in several steps
fixing the explicit solution. Specializing to the ground state sector, we show that our result coincides
with Baxter's conjecture. The Section ends with the discussion of some non-trivial selection rules for the spin matrix
elements.
In Section 4 we sandwich the expression for the spin operator between eigenvectors of the Hamiltonian.
The sum over the intermediate labels is performed explicitly, resulting in fully factorized expressions.
Finally, performing the thermodynamic limit in Section 5, the formula for the order parameters is obtained.
Section 6 summarizes our results. Technical details about performing the factorization and the thermodynamic limit
are given in Appendices, together with a numerical example.

\section{The superintegrable chiral Potts quantum chain}

\subsection{The Hamiltonian of the superintegrable chiral Potts  quantum chain}

The superintegrable  {${\mathbb Z}_N$-symmetric chiral}  Potts  quantum chain  of length $L$
 is defined by the Hamiltonian \cite{GR}
\be\label{H}
\CH\:= \:\CH_0\:+\:\ks\:\CH_1\,,
\ee
where    $\;{\sf k'}$ is a temperature-like real parameter,
    $\;\omega\:=\:e^{2\pi {\rm i}/N},\;\;N\in{\mathbb Z},\;\; N\ge 2\,,\;$ and
\be \label{H0}
\CH_0\:=\:-2\sum_{j=1}^L\:\sum_{n=1}^{N-1} \frac{Z_j^n\: Z_{j+1}^{-n}}{1-\omega^{-n}}\,,\qquad
\CH_1\:=\:-2\sum_{j=1}^L\:\sum_{n=1}^{N-1} \frac{X_j^n}{1-\omega^{-n}}\,.\ee
The space of states is the $L$-fold tensor product of spaces ${\mathbb C}^N$ at each
site $j$. The operators $Z_j,\;X_j\:\in \mbox{End}\,(\mathbb C^N)^{\otimes L}\:$
act non-trivially only at site $j$:
\[ Z_j\:=\:{\mathbf 1}\;\otimes\ldots\otimes\:{\mathbf 1}\:\otimes
         \underbrace{\:Z\:}_{j-th}\otimes\:{\mathbf 1}\:\otimes\ldots\:\otimes\:{\mathbf 1}, \]
\[ X_j\:=\:{\mathbf 1}\;\otimes\ldots\otimes\:{\mathbf 1}\:\otimes
         \underbrace{\:X\:}_{j-th}\otimes\:{\mathbf 1}\:\otimes\ldots\:\otimes\:{\mathbf 1}. \]
$Z$ and $X$ are the $\ZN$-generalizations of the ${\mathbb Z}_2$-Pauli operators $\sigma_z$
and $\sigma_x$. They satisfy the Weyl relation
$\:Z\,X\:=\:\omega\,X\,Z\;$ and $\;Z^N\:=\:X^N\:=\,{\mathbf 1}$.
We always take periodic boundary conditions $\;Z_{L+1}\:=\:Z_1$. A convenient representation
is to label the state vectors by spin variables $\sigma_j \in \ZN,\;j=1\ldots L$, writing
the state vectors as $\;|\sigma_1,\:\sigma_2,\ldots,\sigma_L\rangle\;\:$ and
\be (Z_j)_{\sigma_j,\sigma'_j}= \delta_{\sigma_j,\sigma'_j} \omega ^{\sigma_j}\,,
\qquad
(X_j)_{\sigma_j,\sigma'_j}= \delta_{\sigma_j,\sigma'_j+1}\,.
\label{ZX} \ee
Both $\;\CH_0\;$ and  $\;\CH_1\;$
commute with the spin rotation operator $R\:=\:X_1\: X_2\, \ldots\, X_L$.
Since $R^N\:=\:1$, the space of states of \SIQC\ decomposes into sectors of fixed
of $\ZN$-charge $Q$ where $\:Q=0,1,\ldots, N-1$, corresponding to the eigenvalues $\omega^Q$ of
the $R$.

It turns out that $0\le {\sf k}' <0.902\ldots$ is the ferromagnetic regime \cite{AMCP-level}, where in the thermodynamic limit
the system acquires a non-zero value of the order parameters. One aim of this paper is to
derive its value as a function of ${\sf k}'$.

The Hamiltonian \r{H} can be derived as a logarithmic derivative with respect to a rapidity parameter
from the homogeneous 2D superintegrable chiral Potts model.
However, we shall avoid using properties of the two-dimensional model.

\subsection{The Onsager algebra and the superintegrable chiral Potts quantum chain}

In the beginning of the 1980'th chiral Potts quantum chains received much attention since for certain
parameter values they exhibit an incommensurate phase. Since no integrable chain of this class was known,
Howes, Kadanoff and den Nijs \cite{HKdN} studied perturbative expansions. They found strong evidence that
the ${\mathbb Z}_3$ version of (\ref{H}) has a terminating low-temperature expansion for the mass gap. Subsequent
numerical studies in \cite{GR} showed: if in the $\ZN$-symmetric
quantum chain the coefficients of the
terms involving $Z_j\,Z_{j+1}$ and $X_j$ are chosen as in (\ref{H0}), many gaps become linear.
This feature is known from the Ising model and motivated further studies of the Hamiltonian \r{H}, \r{H0}.
Now is was known that Hamiltonians of the form $\;\CH\,\sim\:A_0\:+\:\ks\:A_1\;$ are
integrable if $\,A_0\,$ and $\,A_1\,$ satisfy the Dolan-Grady-relations \cite{DG}
\be  \left[A_j,\left[A_j,\left[A_j,\;A_{1-j}\:\right]\right]\right]\;=
             \;16\:\left[A_j,A_{1-j}\,\right],\;\;\;\;\;j=0,\,1. \label{DG} \ee
Indeed, in \cite{GR} it was shown that, putting
\[ A_0=-2{\cal H}_1/N\,,\qquad A_1=2{\cal H}_0/N\,. \]
the relations (\ref{DG}) are satisfied. A pair of operators $A_0,\;A_1$ satisfying (\ref{DG}) generates
a basis of Onsager's infinite dimensional Lie-algebra with the operators $\;A_m,\;m\in{\mathbb Z}$,
$\;\;$$\;G_n,\;n\in{\mathbb N}:$
\[   [A_m,A_n]\:=\:4 G_{m-n}\,,\]
\be [G_m,A_n]= 2A_{m+n}-2A_{n-m}\,,\qquad [G_m,G_n]=0\,.\ee
The study of the representation theory of the Onsager algebra has been started by
B.~Davies \cite{Davies}.
He showed that certain finite-dimensional representations of the Onsager algebra
are related to representations
of a direct sum of several copies of the Lie algebra $sl(2)$, see also \cite{Roan91,DateRoan,Roan}.
In the case of the \SIQC\ only two-dimensional representations of $sl(2)$ appear.
Unlike the $N=2$ Ising case, for $N\ge 3$
the Onsager algebra alone is not powerful enough to fully determine the eigenvalues.
In each Onsager  sector of the \SIQC\ the $2^{m_E}$ eigenvalues of $\cal H$ have the Ising-like form
\be
E=A+{\sf k}' B- N \sum_{j=1}^{m_E} \pm \ve(\theta_j)\,,
\ee
with
\be\ve(\theta)=\sqrt{1-2{\sf k}' \cos \theta +{\sf k}'^2}\,.  \label{epsilon}\ee
The constants $A$ and $B$ and the $m_E$ values $\theta_j$ label irreducible
representations of the Onsager algebra.
E.g. for $N=3$ values of $m_E$ can be found in \cite{DKMC}.
However, detailed formulas for $m_E$ will not be used in the following.

Using functional relations which follow from the general integrable 2-dimensional chiral Potts model \cite{BBP},
or, recently, exploiting the loop symmetry of the related $\tau_2$-model \cite{NDCP}, the $\theta_j$ can be calculated
from the zeros of the Baxter-Albertini-McCoy-Perk (BAMP) polynomials \cite{BaxFE,AdvSt,BaxSkew}.
 However, here we shall follow the
spirit of Baxter's recent work and determine the spin-operator matrix elements from the Onsager algebra
and an extension of this algebra alone. So, the $\theta_j$ will be considered as free parameters.
Of course, specific values of the $\theta_j$ calculated from the zeros of the BAMP-polynomials may be inserted in
the final formulas.

\subsection{Extension of the Onsager algebra by a spin operator}

Baxter proposed \cite{Bax3} to extend the Onsager algebra of the \SIQC\ by the local spin operator
$S=Z^r_1$ where $r$ may be chosen to take the values $r=1,\ldots N-1$.
He introduced two operations on operators $X$ in the space of states
\be\label{f0f1}
f_0(X)=\frac{[{\cal H}_0,X]}{2N}\,,\qquad \mbox{and}\qquad
f_1(X)=\frac{[{\cal H}_1,X]+2rX}{2N}
\ee
and has shown that
\be\label{f0f1C}
f_0(S)=0\,, \qquad \mbox{and}\qquad
f_1(f_1(S))=f_1(S)\,.
\ee
Repeated commutation of $S$ with ${\cal H}_0$ and ${\cal H}_1$
produces many new elements. However, Baxter made the conjecture that in the case when
the finite-dimensional representations of Onsager algebra are related to two-dimensional
representations of $sl(2)$ (in particular, in the case of the \SIQC) already the mentioned two
relations for $S$ uniquely
(up to a multiplier because the relations are homogenous) define the matrix elements of $S$.
In the following section we prove this conjecture.

\section{Conjecture on the reduced matrix $S_{PQ}$: a generalization, reformulation and a proof}

\subsection{Generalization}\label{Gen}

We consider the restriction of Hamiltonian $\cal H$ on the subspaces which are spaces of irreducible
representations of the Onsager algebra. The Hamiltonian $\cal H$ on these spaces has an Ising-like spectrum.
We call these spaces {\em Onsager sectors}.
Let ${\cal V}^P$, ${\cal V}^Q$ be some Onsager sectors with $\ZN$-charges $P$, $Q$
and dimensions $2^{m_E^P}$, $2^{m_E^Q}$, respectively.
We will use the short notations $m=m_E^P$, $n=m_E^Q$.

We denote $r=Q-P$ for $P<Q$, and $r=Q-P+N$ for  $P>Q$.
For the sector ${\cal V}^P$ we have
\[
H^P=H_0^P+{\sf k}'H_1^P\,,\qquad H_0^P=A^P-NF^P_0\,,\qquad H_1^P=B^P+N F^P_1\,,
\]
where $A^P$ and $B^P$ are some constants depending on this Onsager sector and
\[
F^P_0=\sum_{j=1}^{m}\left(\begin{array}{cc} 1 & 0\\ 0 & -1\end{array}\right)_j\,,\qquad
F^P_1=\sum_{j=1}^{m}\left(\begin{array}{cc} \cos \theta_j & \sin \theta_j\\ \sin \theta_j & -\cos \theta_j\end{array}\right)_j\,.
\]
We write the same formulas for the sector with charge $Q$ and with parameters $\theta'_j$, $j=1,2,\ldots, n$:
\[
F^Q_0=\sum_{j=1}^{n}\left(\begin{array}{cc} 1 & 0\\ 0 & -1\end{array}\right)_j\,,\qquad
F^Q_1=\sum_{j=1}^{n}\left(\begin{array}{cc} \cos \theta'_j & \sin \theta'_j\\ \sin \theta'_j & -\cos \theta'_j\end{array}\right)_j\,.
\]
Using the fact that ${\cal H}_0$ and ${\cal H}_1$ are block-diagonal with respect to irreducible representations
of the Onsager algebra and
taking  the formulas \r{f0f1} between arbitrary states from ${\cal V}^P$ and ${\cal V}^Q$, we obtain
\be\label{f0def}
f_0(X_{PQ})=-\frac{1}{2} (F_0^P X_{PQ}-X_{PQ} F_0^Q+\alpha X_{PQ})\,,
\ee\be\label{f1def}
f_1(X_{PQ})=\frac{1}{2} (F_1^P X_{PQ}-X_{PQ} F_1^Q+(\beta+1)X_{PQ})\,,
\ee
where $X_{PQ}$ is rectangular submatrix in $X$ and
\be\label{abAB}
\alpha=(A^Q-A^P)/N\,,\qquad
\beta=(B^P-B^Q+2r-N)/N\,.
\ee
Therefore it follows from the relations \r{f0f1C} that the
rectangular submatrix $S_{PQ}$ of the size $2^{m}\times 2^{n}$
of spin operator $S=Z_1^r$ satisfies the system of equations
\be\label{F0C}
F_0^P C-C F_0^Q+\alpha C=0\,,
\ee\be\label{F1C}
F_1^P F_1^P C -2F_1^P C F_1^Q + C F_1^Q F_1^Q+ 2\beta (F_1^P C-C F_1^Q) + (\beta^2-1) C=0
\ee
with respect to the unknown matrix $C$.
The $F_0^P$, $F_0^Q$, $F_1^P$, $F_1^Q$ are given above and
$\alpha$, $\beta$ are supposed to be arbitrary constants.
Formulas \r{f0def}, \r{f1def}, \r{F0C}, \r{F1C} generalize Baxter's relations from
\cite{Bax3} which were written only for ground state Onsager sectors.

The system of equations \r{F0C}, \r{F1C} is homogeneous and any solution of this system multiplied by a constant is
also a solution.
We will find a solution $C_{PQ}$ of the system \r{F0C}, \r{F1C}
and show that space of solutions is no more than one-dimensional.
Therefore {\em exact}\footnote{In the following the term ``exact matrix elements'' will be used in the sense
that the constants $\:{\cal N}_{PQ}\:$ have been determined too.}
 matrix elements $S_{PQ}$ of the spin operator $S=Z_1^r$ are related to this solution
by relation $\:S_{PQ}={\cal N}_{PQ}\; C_{PQ}$ with an unknown scalar multiplier $\:{\cal N}_{PQ}$,
which should be fixed from other arguments.
The other relations of the extended Onsager algebra do not spoil the equality $S_{PQ}={\cal N}_{PQ} C_{PQ}$, but they may impose that
${\cal N}_{PQ}=0$.

\subsection{Reformulation}

Multiplying the relations \r{F0C}, \r{F1C} from the right by
\be\label{Gmat}
G=\prod_{j=1}^{n} \sigma_j^{\rm x}\sigma_j^{\rm z} \,,
\ee
using $F_0^Q=-G^{-1} F_0^QG$, $F_1^Q=-G^{-1} F_1^QG$  and denoting $\tilde C=CG$, we get
similar relations but with changed signs at $F_0^Q$ and $F_1^Q$:
\be\label{F0Cp}
F_0^P \tilde C+\tilde C F_0^Q+\alpha \tilde C=0\,,
\ee
\be\label{F1Cp}
F_1^P F_1^P \tilde C +2F_1^P \tilde C F_1^Q + \tilde C F_1^Q F_1^Q+ 2\beta (F_1^P \tilde C+\tilde C F_1^Q) + (\beta^2-1) \tilde C=0\,.
\ee

Let us combine the matrix elements $\tilde C_{\gamma,\beta}$ into a vector
$\tilde K \in {\cal V}^P \otimes{\cal V}^Q$  with components $\tilde K_{(\gamma,\beta)}$ labeled by multi-indices.
The vector $\tilde K$ has dimension $2^\mu$, where $\mu=m+n$.
Let for $\;k=0,1$
\[
F_k=F_k^P\otimes {\bf 1} + {\bf 1} \otimes (F_k^Q)^t
\]
be an operator in ${\cal V}^P \otimes{\cal V}^Q$ and $(F_k^Q)^t$ is the transposition of $F_k^Q$.
Then the relations \r{F0Cp} and \r{F1Cp} become
\[
(F_0+\alpha)\tilde K=0\,,\qquad (F_1+\beta-1) (F_1+\beta+1)\tilde K=0\,.
\]
This procedure puts the parameters $\{\theta_j\}$ and $\{\theta'_j\}$
on an equal footing: we combine these two sets into the set
$\{\vtheta_1,\ldots,\vtheta_\mu\}=\{\theta_1,\ldots,\theta_{m},\theta'_1,\ldots,\theta'_{n}\}$.

Since the matrices $F_0$ and $F_1$ are both traceless with determinant $-1$
we can find a similarity transformation $F_1=\tilde U^{-1}\,F_0\: \tilde U$ to get
\[
(F_0+\beta-1) (F_0+\beta+1)\tilde U \tilde K=0\,,
\]
where
\be\label{Ud}
\tilde U= U U_d\,,\quad U =\prod_{j=1}^{\mu}\left(\begin{array}{cc} \cos \vtheta_j+1 & \cos \vtheta_j-1  \\ 1 & 1\end{array}\right)_j
\,,\quad
U_d= \prod_{j=1}^{\mu}\left(\begin{array}{cc} 1 & 0  \\ 0 & -\cot \vtheta_j/2\end{array}\right)_j \,.
\ee
It gives finally two relations for the vector $K=U_d \tilde K$:
\be\label{F01}
(F_0+\alpha)K=0\,,\qquad (F_0+\beta-1) (F_0+\beta+1)UK=0\,,
\ee
where we use the fact that $F_0$ commutes with $U_d$. In the next two subsections
we will solve the system of equations \r{F01} with respect to $K$ for arbitrary parameters $\alpha$, $\beta$.

\subsection{Number of solutions of (\ref{F01})}

The relations \r{F01} are a system of linear homogeneous equations with respect to the
unknown components of $K$.
In this subsection we estimate the number of solutions of this system,
depending on $\alpha$ and $\beta$. We use the notation $F=F_0$.

\noindent
{\bf Theorem 1.}  {\it  The dimension of the space of solutions $K$ of the system
\be\label{frel}
(F+\alpha)K=0\,,
\ee\be\label{srel}
(F+\beta-1) (F+\beta+1)UK=0\,,
\ee
where
\be\label{FU}
F=\sum_{j=1}^{\mu}\left(\begin{array}{cc} 1 & 0\\ 0 & -1\end{array}\right)_j\,,\qquad
U =\prod_{j=1}^{\mu}\left(\begin{array}{cc} c_j+1 & c_j-1  \\ 1 & 1\end{array}\right)_j
\ee
at generic values of the parameters $c_k=\cos\vtheta_k$, $k=1,\ldots,\mu$, is no more than one-dimensional.
It is one-dimensional if and only if

\smallskip

\be\label{mueven}
\hspace{-4cm} \bullet \quad \mu\quad \mbox{is even},\quad  \alpha=0, \quad \beta=\pm 1,
\ee
\be\label{muodd}
\hspace{-4cm} \bullet \quad \mu \quad \mbox{is odd},\quad  \alpha=\pm 1,\quad \beta=0.
\ee
}

\medskip

\noindent
{\bf Proof.} The spectrum of $F$ is $\mu,\mu-2,\ldots,-\mu$. From the first relation \r{frel}
we have a necessary restriction on $\alpha$ in order to have $K\ne 0$:
{\it $\alpha$ should be an integer of the same parity as $\mu$, and $|\alpha|\le \mu$}. If $\alpha$ satisfies this condition,
the space of solutions of \r{frel} has dimension $\left(\begin{array}{c} \mu\\ (\mu +\alpha)/2\end{array}\right)$.
The second relation \r{srel} requires the existence of such $K$ from this subspace which are annihilated by the operator
$(F+\beta-1) (F+\beta+1)U$.
Thus, to have $K\ne 0$,  {\it $\;\:\mu+\beta\:$ should be an odd integer and $\:|\beta|\le \mu+1$}. Otherwise, since $U$ is invertible,
we would have that $(F+\beta-1) (F+\beta+1)U$ is also invertible and therefore $K=0$.

We will prove the theorem by induction over $\mu$. Let $F=F_{(\mu)}$ be the initial matrix. In a basis in which the
$\mu$-th tensor component is separated, we have
\[F_{(\mu)}+\alpha=F_{(\mu-1)}+\left(\begin{array}{cc} 1 & 0\\ 0 & -1\end{array}\right)_\mu+\alpha=
\left(\begin{array}{cc} F_{(\mu-1)}+\alpha+1 & 0\\ 0 & F_{(\mu-1)}+\alpha-1\end{array}\right)\,.
\]
It is a diagonal matrix and the zeros on its diagonal define
the space of solutions of \r{frel}. Let ${\cal P}_{(\mu)}(\alpha)$ be the projector to this subspace.
Its rank is \[
\mathop{\rm rk}\, {\cal P}_{(\mu)}(\alpha) =\left(\begin{array}{c} \mu\\ (\mu +\alpha)/2\end{array}\right)\,.
\]
In the basis with separated $\mu$-th tensor component this projector is
\[
{\cal P}_{(\mu)}(\alpha)= \left(\begin{array}{cc} {\cal P}_{(\mu-1)}(\alpha+1) & 0\\
    0 & {\cal P}_{(\mu-1)}(\alpha-1)\end{array}\right)\,.
\]
We need to calculate the rank
\[
f(\mu,\alpha,\beta)=\mathop{\rm rk}\,
\left((F_{(\mu)}+\beta-1) (F_{(\mu)}+\beta+1)U_{(\mu)} {\cal P}_{(\mu)}(\alpha)\right)\,,
\]
since the difference
\[
\tilde f (\mu,\alpha,\beta)= \mathop{\rm rk}\, {\cal P}_{(\mu)}(\alpha)-f(\mu,\alpha,\beta)\ge 0\,
\]
defines the dimension of the space of solutions of the system of equations \r{frel} and \r{srel}.

Extracting explicitly the $\mu$-th tensor component we have
\[
f(\mu,\alpha,\beta)\;=\;{\rm rk}\left(\begin{array}{cc} A & B\\ C & D\end{array}\right)\,,
\]
where
\[
A=(c_\mu+1)(F_{(\mu-1)}+\beta+2)(F_{(\mu-1)}+\beta)U_{(\mu-1)}{\cal P}_{(\mu-1)}(\alpha+1)\,,
\]\[\
B=(c_\mu-1)(F_{(\mu-1)}+\beta+2)(F_{(\mu-1)}+\beta)U_{(\mu-1)}{\cal P}_{(\mu-1)}(\alpha-1)\,,
\]\[
C=(F_{(\mu-1)}+\beta)(F_{(\mu-1)}+\beta-2)U_{(\mu-1)}{\cal P}_{(\mu-1)}(\alpha+1)\,,
\]\[
D=(F_{(\mu-1)}+\beta)(F_{(\mu-1)}+\beta-2)U_{(\mu-1)}{\cal P}_{(\mu-1)}(\alpha-1)\,.
\]
By $f(\mu,\alpha,\beta)$ we denote the rank which is calculated at
generic  values of the parameters $c_k$, $k=1,\ldots,\mu$.
At some specific values of these parameters,
the corresponding rank may be smaller than at generic  values.
At $c_\mu=1$ we have $B=0$ and
using the fact that the rank of a block-triangular matrix is greater or equal
to the sum of ranks of diagonal blocks we obtain
\be\label{fAD}
f(\mu,\alpha,\beta)\ge {\rm rk}\,A+{\rm rk}\,D=f(\mu-1,\alpha+1,\beta+1)+f(\mu-1,\alpha-1,\beta-1)\,.
\ee
Similarly, at $\;c_\mu=-1$ we have $A=0$ and
\be\label{fBC}
f(\mu,\alpha,\beta)\ge {\rm rk}\,B+{\rm rk}\,C=f(\mu-1,\alpha-1,\beta+1)+f(\mu-1,\alpha+1,\beta-1)\,.
\ee
Using identities between binomial coefficients,
the inequalities \r{fAD} and \r{fBC} become
\be\label{ftAD}
\tilde f(\mu,\alpha,\beta)\le \tilde f(\mu-1,\alpha+1,\beta+1)+ \tilde f(\mu-1,\alpha-1,\beta-1)\,,
\ee\be\label{ftBC}
\tilde f(\mu,\alpha,\beta)\le \tilde f(\mu-1,\alpha-1,\beta+1)+\tilde f(\mu-1,\alpha+1,\beta-1)\,.
\ee
We want to show by induction in $\mu$ that
\be\label{fineq}
\tilde f(\mu,\alpha,\beta)\le \delta_{\alpha^2+\beta^2, 1}\,.
\ee
After proving this inequality for $\mu-1$ and all $\alpha$, $\beta$,
a straightforward analysis shows that the inequalities \r{ftAD}, \r{ftBC} imply \r{fineq}.

Assuming that the binomial coefficients $\left(\begin{array}{c} k\\ l\end{array}\right)$ are defined to be $0$
if $l<0$ or $l>k$, we can start the induction from the case $\mu=1$ and arbitrary $\alpha$, $\beta$.
In this case, as it has been explained in the beginning of the proof, in order to have $K\ne 0$ we need to choose
$\alpha=\pm 1$ and $\beta\in \{-2,0,2\}$.
Let for definiteness $\alpha=1$. Then
\[
f(1,\alpha,\beta)=\mathop{\rm rk} \,\left((F_{(1)}+\beta-1) (F_{(1)}+\beta+1)U_{(1)} {\cal P}_{(1)}(\alpha)\right)=
\]\[
\qquad =\left(\begin{array}{cc} (\beta+2)\beta & 0\\ 0 & \beta(\beta-2)\end{array}\right)
U \left(\begin{array}{cc} 0 & 0\\ 0 & 1\end{array}\right)\,.
\]
If $\beta=0$ then $f(1,\alpha,\beta)=0$, otherwise $f(1,\alpha,\beta)=1$.
Thus only if $\beta=0$ we have $\tilde f(1,\alpha,\beta)=1$.

This proves that the space of solutions $K$ is no more than one-dimensional
if $\alpha$ and $\beta$ satisfy the relations \r{mueven} or \r{muodd}. Otherwise there is only the trivial solution $K=0$.
In the next subsection we give the explicit solution for $K$ if
$\alpha$ and $\beta$ satisfy \r{mueven} or \r{muodd}. It proves that in these cases the space of solutions
is one-dimensional. \hfill $\Box$

\medskip

What is the condition on the parameters $c_k$, $k=1,\ldots,\mu$, which leads the theorem to fail?
The analysis for $\mu\le 3$ gives that this condition is: either two of $c_k$ coincide and equal to $\pm 1$ or
three of $c_k$ are pairwise coinciding. We conjecture that for general $\mu$, the corresponding condition is
too restrictive and that the sets $c_k$ arising from {\SIQC} do not satisfy this condition.

\subsection{The solution for the matrix elements}

We proved that if a non-trivial solution $K$ exists, then it is unique up to a multiplier.
Such a solution exists if and only if $\mu$ is even, $\alpha=0$, $\beta=\pm 1$, or if
$\mu$ is odd, $\alpha=\pm 1$, $\beta=0$.

We now present the solution $K$ in the basis of eigenvectors of $F$.
Each basis vector is a tensor product of the two-dimensional vectors associated with summands of $F$ in \r{FU}.
We label  each basis vector by the subset $I$ of the set $\{1,2,\ldots,\mu\}\,$,
such that the elements of $I$ label the tensor components where
the two-dimensional basis vectors with the eigenvalue $+1$ are located.
The complementary subset $\bar I$ contains
the labels of the tensor components where the two-dimensional basis vectors with the eigenvalue $-1$ are placed.
By $|I|$ we denote the size of the subset $I$.

We use a short notation: the product of an expression containing the variable $c$ indexed by a subset $I$
means the product of this expression over the elements of the subset
with variable $c$ indexed by the elements of $I$.

\medskip

\noindent {\bf Theorem 2.} {\it
Let the parameters $\alpha$ and $\beta$ satisfy
\r{mueven} or \r{muodd}.
The vector $K$ with the components
\be\label{KI}
K_I=\frac{\delta_{|I|,(\mu-\alpha)/2}}{\prod (c_I-c_{\bar I})\;\prod (c_I+1)^\sigma\;\prod (c_{\bar I}-1)^\tau}\,
\ee
is the unique (up to a scalar multiplier) solution of the system
\r{frel} and \r{srel}.
Here
\be\label{sigtau}
\sigma=(\beta-\alpha+1)/2,\qquad \tau=(\beta+\alpha+1)/2\,.
\ee
}

\noindent {\bf Remark.} Due to the special values of $\alpha$ and $\beta$,
each of $\sigma$ and $\tau$ can be either $0$ or $1$.

\medskip

\noindent {\bf Proof.} The numerator $\delta_{|I|,(\mu-\alpha)/2}$ ensures that $K$ satisfies \r{frel}.

Let us prove that $K$ satisfies \r{srel}.
In fact we need to prove that the non-zero components of $M=UK$ are in the eigenspaces
of $F$ with eigenvalues $-\beta-1$ and $-\beta+1$.
The components of the vector $M=UK$ are
\be\label{MJsum}
M_J=\sum_I U_{J,I} K_I=\sum_I \frac{\prod (c_{I\cap J}+1)\; \prod (c_{\bar I\cap J}-1) \delta_{|I|,(\mu-\alpha)/2}}
{\prod (c_I-c_{\bar I})\;\prod (c_{I}+1)^\sigma\;\prod (c_{\bar I}-1)^\tau}\,,
\ee
where we used $\; U_{J,I}=\prod (c_{I\cap J}+1)\; \prod (c_{\bar I\cap J}-1)\;$ which follows from \r{FU}.
The components of $M$ are
\be\label{MJprod}
M_J=\frac{P}{\prod (c_{J}-c_{\bar J})\;\prod (c_{\bar J}+1)^\sigma\;\prod (c_{\bar J}-1)^\tau}
\ee
with an unknown polynomial $P$ in the numerator.
Let us explain the origin of the different factors in the denominator of \r{MJprod}.
First, there are no factors like $\:c_j+1$, $j\in J\:$ in the denominator of
\r{MJprod}, because all such factors with $\:j\in I\:$  in the denominator of \r{MJsum} are cancelled.
Similarly, there are no factors like $c_j-1$, $j\in J$.
Second, there are no factors like $c_j-c_{j'}$, $j,j'\in J$, in the denominator of \r{MJprod}.
Potentially such factors can arise from summands in \r{MJsum} for which $j\in I$ and $j'\in \bar I$. But to each
such summand there corresponds a summand with $I$ replaced by
$\tilde I=\{j'\}\cup I\backslash \{j\}$. A straightforward analysis shows that the sum of these two summands
does not contain a pole at $c_j=c_{j'}$. Similarly there are no factors
like $\:c_j-c_{j'}$, $j,j'\in \bar J\:$ in the denominator of \r{MJprod}.

Let us estimate the degree of $P$ with respect to the variables $c_j$, $j\in J$. It is
\[
\max (1-\sigma-|\bar I|, 1-\tau-|I|)+|\bar J|\ge \deg P\ge 0\,.
\]
The two arguments of the maximum correspond to the fact that each element of $J$ is either in $I$ or in $\bar I$.
The term $|\bar J|$ corresponds to the denominator of \r{MJprod}.
Similarly, the estimate of the degree of $P$ with respect to the variables $c_j$, $j\in {\bar J}$, gives
\[
\max (-\sigma-|\bar I|, -\tau-|I|)+|J|+\sigma+\tau\ge \deg P\ge 0\,.
\]
In order to get $M=UK\ne 0$ both inequalities have to be satisfied. Using $|I|=(\mu-\alpha)/2$, $|\bar I|=(\mu+\alpha)/2$, these give
\[
|\bar J|\ge (\mu+\beta-1)/2\,,\qquad  |J|\ge (\mu-\beta-1)/2\,.
\]
Since $|J|+|\bar J|=\mu$ there are only two possibilities:
$|J|= (\mu-\beta-1)/2$, $|\bar J|= (\mu+\beta+1)/2\;$ or
$\;|J|= (\mu-\beta+1)/2$, $|\bar J|= (\mu+\beta-1)/2$. The corresponding eigenvalue $|J|-|\bar J|$ of $F$ is
$-\beta-1$ or $-\beta+1$, respectively.
It means that $K$ satisfies \r{srel}.
The theorem is proved.
\hfill $\Box$

\medskip

In order to obtain matrix elements of $S=Z^r_1$ between the states from Onsager sectors
we need to rewrite the solution $K$ given by \r{KI} in the original notations.

We will label  each basis vector in ${\cal V}^P$ by the subset $V$ of the set $\{1,2,\ldots,m\}$
corresponding to the tensor components where the two-dimensional basis vectors with the eigenvalue $+1$ of
$F^P_0$ are located. The complementary subset $W$ contains
the labels of the tensor components for which the two-dimensional basis vectors with the eigenvalue $-1$ are placed.
In the same way we define the subsets $V'$ and $W'$ of the set $\{1,2,\ldots,n\}$
to label the basis vectors in the space ${\cal V}^Q$.

We also use the set of indices $\{1,2,\ldots,m+n\}$. We will identify
the indices $\{1,2,\ldots,m\}$ with indices of tensor components of ${\cal V}^P$.
The rest indices $\{m+1,m+2,\ldots,m+n\}$ of the set $\{1,2,\ldots,m+n\}$
are identified with indices $\{1,2,\ldots,n\}$ of tensor components of ${\cal V}^Q$ by subtraction $m$.
This identification defines a one to one correspondence between the subsets in $\{1,2,\ldots,m+n\}$
with pairs of subsets from the sets $\{1,2,\ldots,m\}$
and $\{1,2,\ldots,n\}$.

Using the relation $\tilde C=CG$ with $G$ defined in \r{Gmat} and
the relation $K=U_d \tilde K$ with $U_d$ defined in \r{Ud}
we have for the matrix element of $C$
in the basis labeled by $V$ and $V'$
\[
C_{V,V'}=(-1)^{|V'|} \tilde C_{V,W'}= (-1)^{|V'|} \tilde K_{(V,W')}
\]\be\label{CK}
=(-1)^{|W|}  K_I \prod_{i\in W} \tan \theta_i/2 \prod_{i\in V'} \tan \theta'_i/2\,,
\ee
where $I=(V,W')$ is the subset in $\{1,2,\ldots,m+n\}$
defined by the subsets $V$ and $W'$ according to the identification given above.

Note once more, that Theorem~2 defines
the matrix elements $(C_{PQ})_{V,V'}$  by \r{CK}, \r{KI}
as solutions of the system of homogenous equations \r{F0C} and \r{F1C}.
Exact matrix elements $(S_{PQ})_{V,V'}$ of the spin operator
are given through the relation
\be\label{SNC}
(S_{PQ})_{V,V'}={\cal N}_{PQ} (C_{PQ})_{V,V'}
\ee
with an unknown constant ${\cal N}_{PQ}$ which depends on both Onsager sectors.

\subsection{Matrix elements between ground state Onsager sectors}

For the ground state Onsager sectors we have
\be\label{AA}
A^P=L(1-N)+Nm\,,\qquad A^Q=L(1-N)+Nn\,,\ee
\be\label{BB}
B^P=2P+L(1-N)+Nm\,,\qquad B^Q=2Q+L(1-N)+Nn\,.
\ee

It was argued in \cite{Bax3} that
in the {\em ground state} Onsager sectors one has
$\:S_{V,V'}\,=\,1\:$ for $\:W=\emptyset$, $W'=\emptyset$.
In particular this relation defines exact values of the constants
${\cal N}_{PQ}$ for all pairs of ground state Onsager sectors.

Let us consider the case $P<Q$, $m=n$. Using \r{AA}, \r{BB}, \r{abAB} and \r{sigtau}
we obtain $\mu=2m$, $\alpha=0$, $\beta=-1$, $\sigma=\tau=0$.
The components \r{KI} of $K$ are
\be\label{KImm}
K_I=\frac{\delta_{|I|,m}}{\prod (c_I-c_{\bar I})}\,,
\ee
where $\:c_I\:$ is the set $\{\cos \theta_i, \cos \theta'_j\}$, $i\in V$, $j\in W'$
and $c_{\bar I}$ is the set $\{\cos \theta_i, \cos \theta'_j\}$, $i\in W$, $j\in V'$.
Now using \r{CK} and \r{KImm} we get  $C_{V,V'}$.
{}From the relations $S_{V,V'} = C_{V,V'} {\cal N}_{PQ}$ and
$S_{V,V'}=1$ for $W=\emptyset$, $W'=\emptyset\:$
we derive
\be \label{NPQmm}
{\cal N}_{PQ} = \frac {\prod_{i,j=1}^m (c_i-c'_j)}
{\prod_{j=1}^m  \tan \theta'_j/2}\,.
\ee
Thus our solution for $C_{V,V'}$ implies
\[
S_{V,V'}\: =\: \delta_{|W|,|W'|}  \prod_{i\in W} \tan \theta_i/2 \prod_{i\in W'} \cot \theta'_i/2\;\;\;
\]\[
\;\;\;\;\times\; \frac{  \prod_{i\in W,j\in V'}  (c_i-c'_j)\;\prod_{i\in V,j\in W'}  (c_i-c'_j)}
{\prod_{i\in V,j\in W}  (c_i-c_j)\;\prod_{i\in W',j\in V'}  (c'_i-c'_j)}\,,
\]
This formula for $S_{V,V'}$ coincides with the formula (3.45) conjectured by R.~Baxter in
\cite{Bax3} for the case $P<Q$, $m=n$.
It means that Theorem~1 and Theorem~2 prove this conjecture.

Let us consider the case  $P<Q$, $m=n+1$. We have $\mu=2m-1$, $\alpha=-1$, $\beta=0$, $\sigma=1$, $\tau=0$.
In this case the components of $K$ are
\be\label{KImn}
K_I=\frac{\delta_{|I|,m}}{\prod (c_I-c_{\bar I})\prod (c_I+1)}\,.
\ee
Now using \r{CK} and \r{KImn} we get $\;C_{V,V'}= S_{V,V'} {\cal N}^{-1}_{PQ}$, where
\[
S_{V,V'}\:=\:\delta_{|W|,|W'|}\prod_{i\in W} \sin \theta_i \prod_{i\in W'} (\sin\theta'_i)^{-1}\;\;\;
\]\[\;\;\;\;\times\; \frac{\prod_{i\in W,j\in V'}  (c_i-c'_j)\; \prod_{i\in V,j\in W'}  (c_i-c'_j) }
{\prod_{i\in V,j\in W}  (c_i-c_j)\;\prod_{i\in W',j\in V'}  (c'_i-c'_j)}\,,
\]\be\label{NPQmn}
{\cal N}_{PQ}= \frac{ \prod_{i=1}^m (1+c_i) \cdot \prod_{i=1}^m \prod_{j=1}^n(c_i-c'_j)}
{\prod_{i=1}^n  \tan \theta'_i/2}\,.
\ee
This formula for $S_{V,V'}$ is again Baxter's formula in the case $P<Q$, $m=n+1$.

We have also two other cases corresponding to $P>Q$ which can be
considered analogously. In these cases our calculations also prove the Baxter's formula for $S_{V,V'}$.
The first case is $P>Q$, $m=n$ and we have $\mu=2m$, $\alpha=0$, $\beta=1$, $\sigma=\tau=1$,
\be\label{NQPmm}
{\cal N}_{PQ}= (-1)^m { \prod_{i=1}^m (1+c_i) \ \prod_{j=1}^n  \sin \theta'_j \  \prod_{i=1}^m \prod_{j=1}^n(c_i-c'_j)}\,.
\ee
The second case is $P>Q$, $m=n-1$ and  we have $\mu=2n-1$, $\alpha=1$, $\beta=0$, $\sigma=0$, $\tau=1$,
\be\label{NQPmn}
{\cal N}_{PQ}= (-1)^n { \prod_{j=1}^n  \sin \theta'_j \ \prod_{i=1}^m \prod_{j=1}^n(c_i-c'_j)}\,.
\ee

\subsection{Application: selection rules for spin matrix elements}

We proved that the non-trivial solution exists if and only if $\mu$ is even, $\:\alpha=0$, $\beta=\pm 1\,$, or if
$\mu$ is odd, $\;\alpha=\pm 1$, $\beta=0$. This means that only in these cases we may have non-zero matrix elements
$\:S_{PQ}={\cal N}_{PQ} C_{PQ}$
for spin operators. These selection rules are more fine than the selection rule by ${\mathbb Z}_N$-charge.
They follow from \r{abAB}.
For $r=Q-P$ for $P<Q$, and $r=Q-P+N$ for $P>Q$ we have two cases where $\:C_{PQ}\ne 0$:
\begin{itemize}
\item[$\bullet$]
$\mu=m+n$ is even and $A^Q-A^P=0$, $B^P-B^Q=N\pm N-2r$.
\item[$\bullet$]
$\mu=m+n$ is odd and $A^Q-A^P=\pm N$, $B^P-B^Q=N-2r$.
\end{itemize}

To be precise, these selection rules are not complete because the matrix elements
for spin operators between vectors from two Onsager sectors were found up to a common multiplier ${\cal N}_{PQ}$.
For a pair of specific Onsager sectors this multiplier can be zero even if $\:C_{PQ}\ne 0$.
Additional information is needed whether ${\cal N}_{PQ}$ is zero or not.

We have verified these selection rules numerically for $N=3$, $L=3$. In this case they are complete, that is
all ${\cal N}_{PQ}\ne 0$ for the Onsager sectors defined by these selection rules.

\section{Factorized formula for matrix element of spin operators for a finite chain}

\subsection{Matrix elements of spin operators between the eigenvectors of the Hamiltonian}

As has been explained in Section \r{Gen}, the Hamiltonian of the model restricted to the sector with charge $P$ is
\[
H^P=A^P+{\sf k}' B^P-
N \sum_{j=1}^{m}\left(\begin{array}{cc} 1-{\sf k}' \cos \theta_j & -{\sf k}'\sin \theta_j\\
-{\sf k}'\sin \theta_j & -1+{\sf k}'\cos \theta_j\end{array}\right)_j\,.
\]
The eigenvalues of $\:H^P\:$ are
\[
E=A^P+{\sf k}' B^P- N \sum_{j=1}^{m} \pm \ve(\theta_j)\,,
\]
where
\[
\ve(\theta)=\sqrt{1-2{\sf k}' \cos \theta +{\sf k}'^2}\,,\qquad \ve(0)=1-{\sf k}'\,,\qquad \ve(\pi)=1+{\sf k}'\,.
\]
The $-\ve(\theta_j)$ (resp. $+\ve(\theta_j)$) in the sum corresponds to the presence (resp. absence)
of the excitation with rapidity $\theta_j$ because
the change $\ve(\theta_j)$ to $-\ve(\theta_j)$ increases the energy.
The eigenvector of $H^P$ with the lowest energy (that is without excitations) is
\[
\left(\begin{array}{c} a_0(\theta_1)\\ a_1(\theta_1)
\end{array} \right)
\otimes
\left(\begin{array}{c} a_0(\theta_2)\\ a_1(\theta_2)
\end{array} \right)
\otimes \cdots
\otimes
\left(\begin{array}{c} a_0(\theta_m)\\ a_1(\theta_m)
\end{array} \right)
\,,
\]\[
a_0(\theta)=\frac{{\sf k}' \sin\theta}{\sqrt{{{\sf k}'}^2
\sin^2\theta+(1-{\sf k}' \cos\theta- \ve(\theta))^2}}\,,\]
\[
a_1(\theta)=\frac{1-{\sf k}' \cos\theta- \ve(\theta)}
{\sqrt{{{\sf k}'}^2 \sin^2\theta+(1-{\sf k}' \cos\theta- \ve(\theta))^2}}\,,
\]
\[a_0^2(\theta)+a^2_1(\theta)=1\,.
\]
It is useful to express the components $a_k(\theta_j)$, $k=0,1$,  of the eigenvector
through the energies $\ve(\theta)$:
\[
a_0(\theta)=\sqrt\frac{(\ve(\theta)+\ve(0))(\ve(\pi)+\ve(\theta))}{4\ve(\theta)}\,,
\]\[
a_1(\theta)=-\sqrt\frac{(\ve(\theta)-\ve(0))(\ve(\pi)-\ve(\theta))}{4\ve(\theta)}\,.
\]

The formulas for the eigenvectors with some $\theta_j$ excited are obtained from the formula for the non-excited eigenvector by
replacement $\ve(\theta_j)\to -\ve(\theta_j)$ for all such $\theta_j$.
Therefore it is sufficient to consider matrix elements of spin operator only between non-excited eigenvectors from each Onsager sector
and to make the replacement $\ve(\theta_j)\to -\ve(\theta_j)$ for all excited $\theta_j$ in the final formula.
We use the same formulas for the sector with charge $Q$ and with parameters $\theta'_j$, $j=1,2,\ldots, n$.

In each  Onsager sector ${\cal V}^P$ we call the eigenvector of $H$ with the lowest energy (that is without excitations) the
{\em ground state} and denote it $| g.s. \rangle_P$.
The matrix element between the ground states of two Onsager sectors with
charges $P$ and $Q$, $r=(Q-P)\, \mbox{\rm mod}\, N$, is
\[
{}_{P}\langle g.s.| Z_1^{r} | g.s. \rangle _{Q}= \sum_{V,V'}
\prod_{i\in V}  a_{0}(\theta_i) \prod_{i\in W}  a_{1}(\theta_i)
\prod_{i\in V'}  a_{0}(\theta'_i) \prod_{i\in W'}  a_{1}(\theta'_i) (S_{PQ})_{V,V'}\,.
\]
The exact matrix elements $(S_{PQ})_{V,V'}$
are related to the matrix elements $(C_{PQ})_{V,V'}$ defined by \r{CK}, \r{KI} by
the relation \r{SNC}. Then
\bea \lefteqn{ {}_{P}\langle g.s.| Z_1^{r} | g.s. \rangle _{Q}\ny}\\ &&=\;
 {\cal N}_{PQ} \prod_{i=1}^m  a_{0}(\theta_i) \prod_{i=1}^n  a_{1}(\theta'_i)
\;\sum_{V,V'}
\prod_{i\in W}  \left(-\frac{a_{1}(\theta_i)}{a_{0}(\theta_i)}\tan \frac{\theta_i}{2}\right)
\prod_{i\in V'}  \left(\frac{a_{0}(\theta'_i)}{a_{1}(\theta'_i)}\tan \frac{\theta'_i}{2}\right)\,K_I\ny\\
&=&(-1)^n {\cal N}_{PQ}
(2 {\sf k}')^{\sigma (\mu-\alpha)/2 + \tau (\mu+\alpha)/2+(\mu^2-\alpha^2)/4}
\prod_{i=1}^\mu \left(\frac{(\ve_i+\ve_0)(\ve_\pi+\ve_i)}{4\ve_i}\right)^{1/2}\times\ny\\ &&\;\times\;\:
 \sum_{I}\;
\prod_{i\in \bar I}\:  \left( \frac{\ve_i-\ve_0}{\ve_\pi+\ve_i}\right)
\frac{\delta_{|I|,(\mu-\alpha)/2} }{\prod_{i\in I,j\in \bar I} (\ve^2_j-\ve^2_i)
\prod_{i\in I} (\ve^2_\pi-\ve^2_i)^\sigma\prod_{i\in \bar I} (\ve^2_0-\ve^2_i)^\tau}\,,
\label{MEsum}\eea
where we defined
$\ve_i=\ve(\theta_{i})$ for $1\le i\le m$,
$\ve_i=-\ve(\theta'_{i-m})$ for $m<i\le m+n$ and $\ve_0=\ve(0)$, $\ve_\pi=\ve(\pi)$. Also we used
\[
\cos \theta-\cos \theta' = (\ve^2(\theta')-\ve^2(\theta))/(2 {\sf k}')\,,\qquad
\tan \frac \theta 2 = \sqrt\frac{\ve^2(\theta)-\ve^2(0)}{\ve^2(\pi)-\ve^2(\theta)}\,,
\]\[
-\tan \frac{\theta}{2} \cdot \frac{a_1(\theta)}{a_0(\theta)} = \frac{\ve(\theta)-\ve(0)}{\ve(\pi)+\ve(\theta)}\,,\qquad
\tan \frac{\theta}{2} \cdot \frac{a_0(\theta)}{a_1(\theta)} = \frac{-\ve(\theta)-\ve(0)}{\ve(\pi)-\ve(\theta)}\,.\qquad
\]
In order to perform the summation over $I$ in \r{MEsum} we have to discuss the cases of even and odd $\mu$ separately.

\subsection{Summation in the case of even $\mu$}

In this case we have  $\alpha=0$, $\beta=\pm 1$.
Let us consider the case $\beta=-1$. We have $\sigma=\tau=0$.

We use the following summation formula:
\[\sum_I \frac{\delta_{|I|,\mu/2}}{\prod_{a\in I}(z_a+u)\prod_{b\in \bar I}(z_b+v)\prod_{a\in I, b\in \bar I}(z_a^2-z_b^2)} =\]
\[ = \frac{(-1)^{\mu(\mu-2)/8}(u-v)^{\mu/2}}{\prod_{c}(z_c+u)(z_c+v)\prod_{c < s}(z_s+z_c)}\,,\]
which is proved in Appendix A. This formula at $z_i=\ve_i$, $i=1,\ldots, \mu$, and
$u=-\ve_0$, $v=\ve_\pi$ gives
\[ \sum_{I} \frac{\delta_{|I|,\mu/2} }{\prod_{i\in I} (\ve_i-\ve_0) \prod_{i\in \bar I} (\ve_\pi+\ve_i)
\prod_{i\in I,j\in \bar I} (\ve^2_j-\ve^2_i)}
\]\[
=\frac{(-1)^{\mu(\mu-2)/8} (\ve_\pi+\ve_0)^{\mu/2} }
{\prod_{i=1}^\mu (\ve_i-\ve_0) (\ve_\pi+\ve_i)
\prod_{i<j}(\ve_i+\ve_j)}\,,
\]
which we use to make the summation in \r{MEsum}:
\bea \lefteqn{
{}_{P}\langle g.s.| Z_1^{r} | g.s. \rangle _{Q}\:=\ny}\\
&&={\cal N}_{PQ}\;
  \prod_{i=1}^m  \sqrt\frac{(\ve(\theta_i)+\ve(0))}{2\ve(\theta_i)(\ve(\pi)+\ve(\theta_i))}
 \;\prod_{i=1}^n  \sqrt\frac{(\ve(\theta'_i)-\ve(0))}{2\ve(\theta'_i)(\ve(\pi)-\ve(\theta'_i))}\; \times\ny\\
 &&\times\; \frac{(-1)^{(m+n)(m+n-2)/8+n(n+1)/2} (2 {\sf k}')^{(m+n)^2/4}}
{\prod_{i<j}^m (\ve(\theta_i)+\ve(\theta_j)) \prod_{i<j}^n (\ve(\theta'_i)+\ve(\theta'_j))
\prod_{i=1}^m \prod_{j=1}^n (\ve(\theta_i)-\ve(\theta'_j)) }\,.
\label{a0bm1}\eea

Now let us consider the case of matrix elements between ground states of the ground state Onsager sectors
for which $P<Q$ and $m=n=\mu/2$. In this case we can use the expression for ${\cal N}_{PQ}$
from \r{NPQmm} rewritten in terms of $\ve(\theta)$
\[
{\cal N}_{PQ}=(-1)^{mn}(2 {\sf k}')^{-mn} \prod_{j=1}^n  \sqrt\frac{\ve^2(\pi)-\ve^2(\theta'_j)}{\ve^2(\theta'_j)-\ve^2(0)}\cdot
 \prod_{i=1}^m \prod_{j=1}^n (\ve(\theta_i)^2-\ve(\theta'_j)^2)
\]
to get exact matrix elements of spin operator between ground states
\[{}_{P}\langle g.s.| Z_1^{r} | g.s. \rangle _{Q}=
\prod_{i=1}^m \sqrt\frac{(\ve(\theta_i)+\ve(0))}{2\ve(\theta_i) (\ve(\pi)+\ve(\theta_i))}
\prod_{i=1}^m \sqrt\frac{(\ve(\pi)+\ve(\theta'_i))}{2\ve(\theta'_i) (\ve(\theta'_i)+\ve(0))}
\]\be\label{gsMEmm}
\times \frac{ \prod_{i,j=1}^m (\ve(\theta_i)+\ve(\theta'_j)) }
{\prod_{i<j}^m (\ve(\theta_i)+\ve(\theta_j))(\ve(\theta'_i)+\ve(\theta'_j)) }\,.
\ee

In the case of general $m$, $n$ and $\alpha=0$, $\beta=1$, $\sigma=\tau=1$ the analogue of \r{a0bm1} is
\[
{}_{P}\langle g.s.| Z_1^{r} | g.s. \rangle _{Q}={\cal N}_{PQ}
  \prod_{i=1}^m  \frac{1}{(\ve(\pi)-\ve(\theta_i)) \sqrt{2\ve(\theta_i)(\ve(\theta_i)+\ve(0))(\ve(\pi)+\ve(\theta_i))}}\: \times
\] \[
\times \prod_{i=1}^n \frac{1}{ (\ve(\pi)+\ve(\theta'_i))
  \sqrt {2\ve(\theta'_i)(\ve(\theta'_i)-\ve(0))(\ve(\pi)-\ve(\theta'_i))}}\;\times
 \]\be \label{a0bp1}
 \times \frac{(-1)^{(m+n)(m+n-2)/8+n(n+1)/2} (2 {\sf k}')^{m+n+(m+n)^2/4}}
{\prod_{i<j}^m (\ve(\theta_i)+\ve(\theta_j)) \prod_{i<j}^n (\ve(\theta'_i)+\ve(\theta'_j))
\prod_{i=1}^m \prod_{j=1}^n (\ve(\theta_i)-\ve(\theta'_j)) }\,.
\ee
This gives the matrix elements between ground states of the ground state Onsager sectors
for which $P>Q$ and $m=n=\mu/2$ if one uses the corresponding ${\cal N}_{PQ}$ from \r{NQPmm}.
The final formula is analogous to \r{gsMEmm} with additional factor $(-1)^m$ and with
the replacement of the sets $\{\ve(\theta_i)\}\leftrightarrow \{\ve(\theta'_j)\}$.

\subsection{Summation in the case of odd $\mu$}

In the case of odd $\mu$ we have  $\alpha=\pm 1$, $\beta=0$.
Let us consider the case $\alpha=-1$. Then $\sigma=1$, $\tau=0$.

In this case we will use another summation formula from Appendix A:
\[\sum_I \frac{\delta_{|I|,(\mu+1)/2}}{\prod_{a\in I}(z_a+u)(z_a+v)\prod_{a\in I, b\in \bar I}(z_a^2-z_b^2)}\]
\[ = \frac{(-1)^{(\mu+1)(\mu-1)/8}(u+v)^{(\mu-1)/2}}{\prod_{c}(z_c+u)(z_c+v)\prod_{c < s}(z_s+z_c)}.\]
It allows to perform the summation in \r{MEsum} and get
\[
{}_{P}\langle g.s.| Z_1^{r} | g.s. \rangle _{Q}=
\frac{{\cal N}_{PQ}}{(\ve(0)+\ve(\pi))^{1/2}}    \prod_{i=1}^m
\frac{1}{(\ve(\pi)-\ve(\theta_i))} \sqrt\frac{(\ve(\theta_i)+\ve(0))}{2\ve(\theta_i)(\ve(\pi)+\ve(\theta_i))}
\]\[\times  \prod_{i=1}^n  \frac{1}{(\ve(\pi)+\ve(\theta'_i))}
\sqrt\frac{(\ve(\theta'_i)-\ve(0))}{2\ve(\theta'_i)(\ve(\pi)-\ve(\theta'_i))}
\]\be \label{am1b0}
\times
\frac{(-1)^{(\mu^2-1)/8+n(n+1)/2} (2 {\sf k}')^{(\mu+1)^2/4}  }
{\prod_{i<j}^m (\ve(\theta_i)+\ve(\theta_j)) \prod_{i<j}^n (\ve(\theta'_i)+\ve(\theta'_j))
\prod_{i=1}^m \prod_{j=1}^n (\ve(\theta_i)-\ve(\theta'_j)) }\,.
\ee

Now let us consider the case of matrix elements between ground states of the ground state Onsager sectors
for which $P<Q$ and $m=n+1$, $m=(\mu+1)/2$, $n=(\mu-1)/2$. In this case we can use the exact expression for ${\cal N}_{PQ}$
from \r{NPQmn} rewritten in terms of $\ve(\theta)$:
\bea \lefteqn{ {\cal N}_{PQ}\;=\;\ny}\\
&=&\!(-1)^{mn}(2 {\sf k}')^{-mn} \prod_{i=1}^m \prod_{j=1}^n (\ve(\theta_i)^2\!-\ve(\theta'_j)^2)
\prod_{i=1}^m \frac{\ve(\pi)^2-\ve(\theta_i)^2}{2{\sf k}'}
\prod_{j=1}^n \sqrt\frac{\ve^2(\pi)-\ve^2(\theta'_j)}{\ve^2(\theta'_j)-\ve^2(0)}
\ny\eea
to get the exact matrix elements between ground states:
\[{}_{P}\langle g.s.| Z_1^{r} | g.s. \rangle _{Q}=
\]\[
=\prod_{i=1}^m  \sqrt\frac{(\ve(\theta_i)+\ve(0))(\ve(\pi)+\ve(\theta_i))}{2\ve(\theta_i)}
\prod_{i=1}^n \sqrt\frac{1}{2\ve(\theta'_i) (\ve(\theta'_i)+\ve(0))(\ve(\pi)+\ve(\theta'_i))}\;\times
\]\be \label{gsMEmn}
\times
\frac{ \prod_{i=1}^m \prod_{j=1}^n (\ve(\theta_i)+\ve(\theta'_j)) }
{(\ve(\pi)+\ve(0))^{1/2} \prod_{i<j}^m (\ve(\theta_i)+\ve(\theta_j)) \prod_{i<j}^n (\ve(\theta'_i)+\ve(\theta'_j))  }\,.
\ee

In the case of general $m$, $n$ and $\alpha=1$, $\beta=0$, $\sigma=0$, $\tau=1$ the analogue of \r{am1b0} is
\[
{}_{P}\langle g.s.| Z_1^{r} | g.s. \rangle _{Q}=\frac{{\cal N}_{PQ}}{(\ve(0)+\ve(\pi))^{1/2}}
  \prod_{i=1}^m  \frac{1}{\sqrt{2\ve(\theta_i)(\ve(\theta_i)+\ve(0))(\ve(\pi)+\ve(\theta_i))}} \times
\] \[
\times \prod_{i=1}^n \frac{1}{ \sqrt {2\ve(\theta'_i)(\ve(\theta'_i)-\ve(0))(\ve(\pi)-\ve(\theta'_i))}} \times
 \]\be \label{ap1b0}
 \times \frac{(-1)^{(m+n+1)(m+n+3)/8+n(n+1)/2} (2 {\sf k}')^{(m+n+1)^2/4}}
{\prod_{i<j}^m (\ve(\theta_i)+\ve(\theta_j)) \prod_{i<j}^n (\ve(\theta'_i)+\ve(\theta'_j))
\prod_{i=1}^m \prod_{j=1}^n (\ve(\theta_i)-\ve(\theta'_j)) }\,.
\ee
It gives the matrix elements between ground states of the ground state Onsager sectors
for which $P>Q$ and $m=n-1$ if one uses the corresponding ${\cal N}_{PQ}$ from \r{NQPmn}.
The final formula is analogous to \r{gsMEmn} with an additional factor $(-1)^n$ and with
the replacement of the sets $\{\ve(\theta_i)\}\leftrightarrow \{\ve(\theta'_j)\}$.

\section{Order parameters for the superintegrable chiral Potts quantum chain}

\subsection{Thermodynamic limit of ground state matrix elements}

For the superintegrable chiral Potts quantum chain with Hamiltonian \r{H}
we have $N-1$ order parameters ${\cal S}_r$, $r=1,2,\ldots,N-1$,
which take non-zero values in the ferromagnetic phase $0\le {\sf k}'<1$ and
are defined by the matrix elements between ground states of ground state Onsager sectors (labeled by charge $P$)
in the thermodynamic limit:
\be
{\cal S}_r=\mathop{\lim}_{L\to\infty} {}_{P}\langle g.s.| Z_1^{r} | g.s. \rangle _{P+r}\,.
\label{limo}\ee
At  $0\le {\sf k}'<1$ there is $N$-fold asymptotic ($L\to \infty$) degeneration of
energies of these ground states.
(In fact, for $\;0.901292\ldots <{\sf k}'<\:1\:$ these translationally invariant states are not the true
ground states, i.e. states with lowest energies \cite{AMCP-level,AdvSt}. Due to level crossing, the lowest energy state
becomes non-translationally invariant).
In the paper \cite{ACPT}, from perturbative calculations the formula for
the order parameters
\be\label{op} {\cal S}_r=(1-{{\sf k}'}^2)^{r(N-r)/(2N^2)}
\ee
was conjectured. Recently this formula was proved by Baxter \cite{BaxOPL,BaxOP}
using functional relations and analytical properties. Below we derive this formula
starting from the exact results for the matrix elements of spin operator between ground states in
the chain of finite length $L$.

{}From \r{gsMEmm} the square of ground states matrix elements in the case of $\:m=n\:$ is
\be\label{conj}
\begin{array}{ll}
\ds {}_{P}\langle g.s.| Z_1^{r} | g.s. \rangle _{P+r}^2&\ =\ \ds
\frac{\prod_{i=1}^m (\ve(\theta_i)+\ve(0))(\ve(\theta'_i)+\ve(\pi))}
{\prod_{i=1}^m (\ve(\theta_i)+\ve(\pi))(\ve(\theta'_i)+\ve(0))}\\
&\ \times\ \ds \frac{\prod_{i,j=1}^m (\ve(\theta_i)+\ve(\theta'_j))^2}
{\prod_{i,j=1}^m (\ve(\theta_i)+\ve(\theta_j)) (\ve(\theta'_i)+\ve(\theta'_j))}\,,
\end{array}
\ee
where the $\theta_i$ and $\theta'_i$ refer to the Onsager sectors with charges $P$ and $P+r$, respectively.

In the case $\:m=n+1\:$ from \r{gsMEmn} the matrix element of the spin operator $Z_1^{r}$ between the ground states is
\be\label{conjec}
\begin{array}{ll}
\ds {}_{P}\langle g.s.| Z_1^{r} | g.s. \rangle _{P+r}^2&\ =\ \ds
 \frac{\prod_{i=1}^{m} (\ve(\theta_i)+\ve(0))(\ve(\theta_i)+\ve(\pi))}
{(\ve(0)+\ve(\pi))\prod_{i=1}^n (\ve(\theta'_i)+\ve(0))(\ve(\theta'_i)+\ve(\pi))}\\
&\ \times\ \ds \frac{\prod_{i=1}^n\prod_{j=1}^{m}(\ve(\theta'_i)+\ve(\theta_j))^2}
{\prod_{i,j=1}^n (\ve(\theta'_i)+\ve(\theta'_j))\prod_{i,j=1}^{m} (\ve(\theta_i)+\ve(\theta_j))}\,.
\end{array}
\ee

{}From \r{limo} and \r{op} the formula for the order parameters which we have to obtain in the thermodynamic limit is
\be  \label{Ordp}
\mathop{\lim}_{L\to\infty} {}_{P}\langle g.s.| Z_1^{r} | g.s. \rangle _{P+r}^2=(1-{{\sf k}'}^2)^{r(N-r)/N^2}\,.
\ee

First we consider the case of matrix elements of type \r{conj} when $m=n$.
Whereas up to now the $\theta_i$ and $\theta_i'$ were arbitrary, now we take them to be
determined by the BAMP-polynomials of
\cite{BaxFE,AdvSt}.
We use the following relation for the thermodynamic limit:
\be\label{thlimz}\phi (\lm):=
\mathop{\lim}_{L\to\infty}
\left(
\sum_{i=1}^m \log (\lm+\ve(\theta_i))-\sum_{i=1}^m \log (\lm+\ve(\theta'_i))\right)
= \frac{r}{N} \log \frac{\lm+1-{\sf k}'}{\lm+1+{\sf k}'}\,,
\ee
where $\lm$ is arbitrary parameter. The proof of this formula is given in Appendix~B.
The thermodynamic limit is taken in such a way to keep the relation $m=n$ and values of $P$ and $r$.
This limit can be realized by adding multiples of $N$ to the chain length $L$.

Using \r{thlimz} at $\lm=\ve(0)=1-{\sf k}'$ and $\lm=\ve(\pi)=1+{\sf k}'$ we obtain
\be\label{thlim0}
\mathop{\lim}_{L\to\infty}  \frac{\prod_{i=1}^m (\ve(\theta_i)+\ve(0))}
{\prod_{i=1}^m (\ve(\theta'_i)+\ve(0))}= (1-{\sf k}')^{r/N}\,,
\ee\be\label{thlimpi}
\mathop{\lim}_{L\to\infty} \frac{\prod_{i=1}^m (\ve(\theta'_i)+\ve(\pi))}
{\prod_{i=1}^m (\ve(\theta_i)+\ve(\pi))}= (1+{\sf k}')^{r/N}\,.
\ee
The thermodynamic limit of the double products in \r{conj} can be calculated along the
following chain of equalities:
\[
\mathop{\lim}_{L\to\infty} \log \frac{\prod_{i,j=1}^m (\ve(\theta_i)+\ve(\theta'_j))^2}
{\prod_{i,j=1}^m (\ve(\theta_i)+\ve(\theta_j)) (\ve(\theta'_i)+\ve(\theta'_j))}
\]\[
=\mathop{\lim}_{L\to\infty} \left ( \sum_{i,j} \left( \log (\ve(\theta_i)+\ve(\theta'_j))-
\log (\ve(\theta'_i)+\ve(\theta'_j))\right)\right. \]
\[+
\left. \sum_{i,j} \left( \log (\ve(\theta'_i)+\ve(\theta_j))-\log (\ve(\theta_i)+\ve(\theta_j))\right)\right)
\]\[
= \mathop{\lim}_{L\to\infty}  \sum_{j} \left(\phi ( \ve(\theta'_j) ) - \phi ( \ve(\theta_j)) \right)
\]\[
=\mathop{\lim}_{L\to\infty}  \frac{r}{N} \left (
\sum_{j} \log(\ve(\theta'_j)+1-{\sf k}') - \sum_{j} \log(\ve(\theta'_j)+1+{\sf k}')
\right.\]\[
\left. -\sum_j\log(\ve(\theta_j)+1-{\sf k}')
+\sum_j \log(\ve(\theta_j)+1+{\sf k}') \right)
\]\[
=\frac{r}{N}(- \phi (1-{\sf k}') + \phi (1+{\sf k}')) =  -\frac{r^2}{N^2} \log(1-{{\sf k}'}^2)\,.
\]
Exponentiating we get
\be\label{thlimij}
\frac{\prod_{i,j=1}^m (\ve(\theta_i)+\ve(\theta'_j))^2}
{\prod_{i,j=1}^m (\ve(\theta_i)+\ve(\theta_j)) (\ve(\theta'_i)+\ve(\theta'_j))}\to (1-{{\sf k}'}^2)^{-r^2/N^2}\,.
\ee
Combining the latter relation with \r{thlim0} and \r{thlimpi} we obtain \r{op}.

In the case of the matrix elements \r{conjec} we have to use the same arguments but instead of the
formula \r{thlimz} we have to explore the relation
\be\label{thlimz1}
\mathop{\lim}_{L\to\infty}
\left(
\sum_{i=1}^m \log (\lm+\ve(\theta_i))-\sum_{i=1}^{m-1} \log (\lm+\ve(\theta'_i))\right)
= \frac{r}{N} \log \frac{\lm+1-{\sf k}'}{\lm+1+{\sf k}'}+\log (\lm+1+{\sf k}')\,.
\ee
So, in both cases for the order parameter we obtain the famous result \r{Ordp}.

\section{Conclusions}

Starting from Baxter's extension of the
Onsager algebra we have found factorized expressions for the spin operator matrix elements  between the
eigenstates of
Hamiltonians of the finite length superintegrable $\ZN$-symmetric chiral Potts quantum chain
up to a scalar factor  for any pair of the Onsager sectors.

Thus the problem of finding all the matrix elements is
reduced to the calculation of these scalar factors.
Derivation of these factors  will probably require an information
on the Bethe-states of $\tau_2$-model and $sl(2)$-loop algebra symmetries.
Further investigation in this direction is important
since the knowledge of the explicit formulas for the matrix elements will be useful for the
study of the correlation functions of the superintegrable chiral Potts chain.
These scalar factors are known for the matrix elements between the ground
states Onsager sectors and this allows to take the thermodynamics limit using the standard technique
 and to derive the order parameters of the superintegrable
chiral Potts chain.

In the case $N=2$ (the Ising quantum chain in a transverse field)
these scalar factors can be found for all possible pairs of the Onsager
sectors. This allows to obtain
the factorized expression for the spin operator matrix elements in this model found in \cite{BBS2} in
the framework of the method of Separation of Variables.
We will address this problem in our forthcoming paper.

\smallskip

{\bf Note added:} When our paper was ready for publication we learned about
Baxter's new paper \cite{Bax4},
which is a continuation of his previous works \cite{Bax2,Bax3}.
In this paper it is shown that the determinant representation (7.11) from
\cite{Bax2} implies the factorized formulas for $\bigl({\cal M}_r^{(2)}\bigr)^2$, which
in the thermodynamic limit give the order parameters of the \SIQC. Baxter's  formulas for
$\bigl({\cal M}_r^{(2)}\bigr)^2$ coincide with our results \r{conj} and \r{conjec}.

\section*{Acknowledgements}
G.v.G. and S.P. have been supported by the Heisenberg-Landau exchange program HLP-2008.
S.P. has also been supported in part by the RFBR
grant  08-01-00392 and the grant for the Support
of Scientific Schools NSh-3036.2008.2.
The work of N.I.,  V.S. and Yu.T was supported
by the Program of Fundamental Research of the Physics and
Astronomy Division of the NAS of Ukraine, the Ukrainian FRSF grants
28.2/083 and 29.1/028,
by French-Ukrainian program ``Dnipro''  M17-2009 and  the joint project PICS of CNRS
and NAS of Ukraine.

\section*{Appendix A. Proof of the summation formulas}

In the case of even $\mu$ we have the following summation formula:
\[
\sum_I \frac{\delta_{|I|,\mu/2}}{\prod_{a\in I}(z_a+u)\prod_{b\in \bar I}(z_b+v)\prod_{a\in I, b\in \bar I}(z_a^2-z_b^2)} \]
\be\label{sumI}
= \frac{(-1)^{\mu(\mu-2)/8}(u-v)^{\mu/2}}{\prod_{c}(z_c+u)(z_c+v)\prod_{c < s}(z_s+z_c)}.
\ee
We start from the left-hand side. It does not have poles at $z_a=z_b$. Indeed, such a pole may arise when $a\in I$, $b\in \bar I$.
But to each such summand there corresponds summand with $I$ replaced by
$\tilde I=\{b\}\cup I\backslash \{a\}$. A straightforward analysis shows that the sum of these two summands
does not contain a pole at $z_a=z_b$. Therefore the left-hand side can be presented as
\[
\frac{P_\mu(u,v;z_1,z_2,\ldots,z_\mu)}{\prod_{c}(z_c+u)(z_c+v)\prod_{c < s}(z_s+z_c)}\,,
\]
where $P_\mu(u,v;z_1,z_2,\ldots,z_\mu)$ is a polynomial of corresponding variables.
Denote the degree of the rational function in the left hand side of \r{sumI} with respect to the variable
$z_c$ by $d_c$ and estimate it. From the initial sum we have $d_c\le -\mu-1$. From the factorized expression we have
$d_c= \deg_{z_c}P_\mu -\mu-1$. Hence $\deg_{z_c}P_\mu=0$. That is $P_\mu$ does not depend on $z_c$ for all $c$.

To find $P_\mu(u,v)$ let us analyze the pole at $z_1\to -z_2$. Such pole arises if $1\in I$, $2\in \bar I$ or
if $2\in I$, $1\in \bar I$.
Then we separate factors depending on $z_1$ and $z_2$ and denote by $I'$ and $\bar I'$ the sets $I$ and $\bar I$ with $1$ and $2$ excluded.
At $z_1\to -z_2$ the summation is reduced to the summation over $I'$ reducing the calculation of the sum
to the same problem for $\mu-2$.
It gives the recurrent relation
\[
P_\mu(u,v)=(-1)^{(\mu-2)/2} (u-v) P_{\mu-2}(u,v)\,.
\]
Calculating explicitly the sum at $\mu=2$ we get $P_2(u,v)=(u-v)$. Therefore $P_\mu(u,v)=(-1)^{\mu(\mu-2)/8}(u-v)^{\mu/2}$.
It proves the summation formula for even $\mu$.

The summation formula for odd $\mu$
\[\sum_I \frac{\delta_{|I|,(\mu+1)/2}}{\prod_{a\in I}(z_a+u)(z_a+v)\prod_{a\in I, b\in \bar I}(z_a^2-z_b^2)} =\]
\[ = \frac{(-1)^{(\mu+1)(\mu-1)/8}(u+v)^{(\mu-1)/2}}{\prod_{c}(z_c+u)(z_c+v)\prod_{c < s}(z_s+z_c)}\]
can be proved in the same way.

\section*{Appendix B. Proof of \r{thlimz}}

The aim of this Appendix is to prove \r{thlimz} and \r{thlimz1}.
The proof is close to the derivation presented in \cite{Baxter89}.

We use the BAMP polynomial in $e^{Nu}$ of degree $m$
\be\label{PsiP}
\Psi_P(u)=e^{-Pu} \sum_{k=0}^{N-1} \omega^{Pk} \left(\frac{1-e^{Nu}}{1-\omega^{-k} e^u}\right)^L\,.
\ee
In most papers the polynomial \r{PsiP} is written in terms of the variable $z=e^u$ (\cite{BaxFE},
similarly in \cite{AdvSt} etc.). The zeros of the polynomials $\Psi_P$
are known to be all simple and to come on the negative real axis of the variable $z^N$. For the following,
the use of the variable $u$ will be useful:
so, \r{PsiP} has simple zeros at $u=u_1,u_2,\ldots,u_{m}$,
where $u_1,u_2,\ldots,u_{m}$ can be chosen to have imaginary part $\pi/N$.
Each zero $e^{N u_j}$ of this polynomial corresponds to a real value of $\theta_j$,
calculated through the relation
\[
\cos \theta = (1+e^{Nu})/(1-e^{Nu}).
\]
This relation also implies that
${\rm Re}\, u\to -\infty$ corresponds to $\theta=0$, $\ve(0)=1-{\sf k}'$,
${\rm Re}\, u\to +\infty$ corresponds to $\theta=\pi$, $\ve(\pi)=1+{\sf k}'$.

\begin{figure}[width=0.75\textwidth]
\begin{center}
  \includegraphics{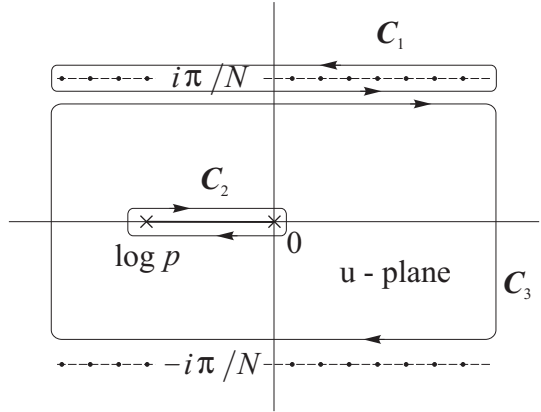}
\caption{Integration contours $C_1,\:C_2,\:C_3$ used in Appendix~B.}
\label{fig:1}
\end{center}
\end{figure}

We present the following sum as an integral along the contour $C_1$ shown in Fig.~1:
\[
\sum_{i=1}^{m} \log (\lm+\ve(\theta_i))= \oint_{C_1} \log \left(\frac{ \ve(\theta)+\lm }{1+{\sf k}'+\lm}\right)
\frac{\Psi'_P(u)}{\Psi_P(u)}  \frac{du}{2\pi{\rm i}}
\]
\be\label{sum_int}
+ {\log (1+{\sf k}'+\lm)} \oint_{C_1}
\frac{\Psi'_P(u)}{\Psi_P(u)}  \frac{du}{2\pi{\rm i}}\,.
\ee
The motivation for introducing the denominator $\log (1+{\sf k}'+\lm)$ in the first term of
r.h.s of \r{sum_int} will be given below.
The integral in the second  term gives $m$. The other sum entering \r{thlimz}, \r{thlimz1}, and
corresponding to sector with charge $P+r$, is given by a similar relation where the second term
is $n\log (1+{\sf k}'+\lm)$. In the case of \r{thlimz} we have $m=n$ and the difference of these terms disappears.
In the case of \r{thlimz1} we have $m=n+1$ and the difference is $\log (1+{\sf k}'+\lm)$.

Let us analyze the first integral  from the right-hand side of \r{sum_int}.
Its integrand is analytic in the domain enclosed by contours $C_1$ and $C_3$ except for the simple poles enclosed
by $C_1$ and a branch cut (where $\ve(\theta)$ is pure imaginary) on the negative
real axis along the segment $u\in [\log p, 0]$ (enclosed by $C_2$), where
\[
\log p=\frac{2}{N}\log \frac{1-{{\sf k}'}}{1+{{\sf k}'}}<0\,.
\]
At $u\to 0$, $\ve(\theta)\to \infty$. At $u \to \log p$, $\ve(\theta)\to 0$.

The integral with the same integrand along the contour $C_1-C_3$ gives $0$. Indeed, the integral
along the horizontal lines vanishes due to the
periodicity of integrand under the transformation $u\to u+2\pi{\rm i}/N$.
The integral along the right vertical segment vanishes due to
\[
\log\left(\frac{ \ve(\theta)+\lm }{1+{{\sf k}'}+\lm}\right)\to 0\,.
\]
This gives the motivation for introducing the denominator ${1+{{\sf k}'}+\lm}$ in the first integral in \r{sum_int}.
The integral along the left vertical segment vanishes
since $\Psi_P(u)$ is a polynomial in $e^{Nu}$ with non-zero free term and therefore
$\Psi'_P(u)/\Psi_P(u)\sim e^{Nu}\to 0$ as ${\rm Re}\, u\to -\infty$.
Therefore instead of the integration along $C_1$ we may integrate along $C_3$.

Now we shrink the contour  $C_3$ to $C_2$, that is around the branch
cut. On  $C_2$, the term in \r{PsiP} with $k=0$ exponentially dominates as $L\to \infty$.
Indeed, the absolute value of the ratio of the term $k=0$ and any other term $k\ne 0$ is the $L$-th power of
\[
\frac{|1-\omega^{-k}e^u|}{1-e^u}>1
\]
with the left-hand side of the inequality minimized at $u=\log p$.
Therefore in thermodynamic limit we can ignore the terms with $k\ne 0$ in the polynomial \r{PsiP} and replace it by
\[
e^{-Pu} [(1-e^{Nu})/(1-e^u)]^L\,
\]
in the integral along $C_2$.
Finally we expand the contour to $C_3$ and get
\[
\frac{1}{2\pi{\rm i}} \oint_{C_3} \log \left(\frac{ \ve(\theta)+\lm }{1+{{\sf k}'}+\lm}\right)
\left(-P+L\left(-\frac{N}{1-e^{-Nu}}+\frac{1}{1-e^{-u}}\right)\right) du\,.
\]
The terms proportional to $L$ dominate at $L\to \infty$ but they do not depend on $P$ and so they disappear in
\r{thlimz} and \r{thlimz1}.
Therefore the left-hand side of \r{thlimz} in the limit $L\to \infty$ becomes
\[
\frac{-P-(-(P+r))}{2\pi{\rm i}} \oint_{C_3} \log \left(\frac{ \ve(\theta)+\lm }{1+{{\sf k}'}+\lm}\right) du=
\]\[
=\frac{r}{2\pi{\rm i}} \log \left(\frac{ 1-{{\sf k}'}+\lm }{1+{{\sf k}'}+\lm}\right) \int_{-\infty-{\rm i}\pi/N}^ {-\infty+{\rm i}\pi/N} du
=\frac{r}{N} \log \left(\frac{ 1-{{\sf k}'}+\lm }{1+{{\sf k}'}+\lm}\right)\,,
\]
where we used the fact that the integral along the horizontal lines vanishes due to the periodicity of the integrand,
and along the right vertical line due to vanishing the integrand. Only the left vertical line gives a constant contribution.
Recall that ${\rm Re}\, u\to -\infty$ corresponds to $\theta=0$, $\ve(0)=1-{{\sf k}'}$.
In the case of  \r{thlimz1} we have the same result but with the additional term $\log (1+{\sf k}'+\lm)$ as was explained above.

\section*{Appendix C. Numerical verification}

Before starting our research we implemented a numerical verification of the
finite-size formula \r{conj}, and compared this to the analytical
formula \r{Ordp} for the square of the order parameter.
We approximate the angles $\;\theta_i\:$ and $\:\theta'_i\:$ for $L\to\infty$ as \cite{G98}
\be \label{theta-approx}
\cos \theta_l=\frac{\sin^N(K_l+\pi/N)-\sin^N(K_l)}{\sin^N(K_l+\pi/N)+\sin^N(K_l)}\,,\ee
\[ \;\;\;\;\;K_l=\frac{2Nl+2Q-N}{2NL}\;\pi\,,
\qquad l=1,2,\ldots,m\,.
\]

\noindent We fix ${\sf k}'=0.8$ and obtain:

\medskip

\noindent $N=3$, $L=300$, $r=1$:\\
Numerical calculation of \r{conj} with \r{theta-approx}:  $|{}_{Q=1}<S>_{Q=2}|^2\:=\:0.796894034089816$\\
Analytic formula: $(1-{{\sf k}'}^2)^{2/9}\:=\: 0.796893997784373$\\
Error: $3.63054*10^{-8}$

\medskip

\noindent $N=5$, $L=300$, $r=1$:\\
Numerical calculation:  $|{}_{Q=1}<S>_{Q=2}|^2\:=\:  0.8491969001045062$\\
Analytic formula: $(1-{{\sf k}'}^2)^{4/25}\:=\: 0.8491969000417509$\\
Error: $6.27554*10^{-11}$

\medskip

\noindent $N=5$, $L=300$, $r=2$:\\
Numerical calculation:  $|{}_{Q=1}<S>_{Q=3}|^2\:=\: 0.7825509089016481$\\
Analytic formula: $(1-{{\sf k}'}^2)^{6/25}\:=\: 0.7825509088837955$\\
Error: $1.07617*10^{-11}$

\medskip
\noindent
These results of the numerical verification encouraged us to
look for an analytical proof of \r{conj}.



\end{document}